\documentclass[12pt]{iopart}
\usepackage{graphicx}
\usepackage{color}

\begin{document}

\title{Optical properties of  Fe-Mn-Ga alloys}

\author{Y. V. Kudryavtsev$^{1}$, N. V. Uvarov$^{1}$ and J. Dubowik$^{2}$}
\address{$^{1}$Institute of Metal Physics, NAS of Ukraine,
Vernadsky 36, 03142 Kiev, Ukraine}
\address{$^{2}$Institute of Molecular Physics, PAS,
Smoluchowskiego 17, 60-179 Pozna\'{n}, Poland}

\begin{abstract}

The first-principles calculations of the electronic structures and
the interband optical conductivity (OC) spectra have been
performed for the stoichiometric Fe$_{2}$MnGa alloy with L2$_{1}$
and L1$_{2}$ types of atomic ordering.  The calculated optical
properties of Fe$_{2}$MnGa alloy for the L2$_{1}$ and L1$_{2}$
phases are complemented by the experimental OC spectra for bulk
and thin film Fe-Mn-Ga alloy samples near the stoichiometry 2:1:1
with L2$_{1}$ and L1$_{2}$ (for bulks) as well as the
body-centered-cubic and face-centered-cubic (for films)
structures, respectively. A reasonable agreement between
experimental and calculated interband OC spectra was obtained for
both phases of the alloy. The experimental data show no
significant difference in the OC spectra with respect to the
degrees of atomic and magnetic orders of the samples.

\end{abstract}


\maketitle

\section{Introduction}

Heusler alloys (HA) attract much attention due to emergent new
physical properties and applications (martensitic transformations,
half-metallicity, metamagnetic transformations etc.) resulting
from their specific electronic structures \cite{graf2011}.

Among various experimental tools for studying the electronic
structures of metals the optical spectroscopy (namely,
spectroscopic ellipsometry) has a great advantage in comparison
with other methods {which provide information on the band
structure} \cite{uba}. Specifically, optical spectroscopy  is
regarded as a method that manifests a high energy resolution of
$\sim$ 0.01 eV within {a} ± 5 - 6 eV energy range near the
Fermi level ($E_{F}$). On the other hand, among various optical
parameters of metals, interband optical conductivity (IBOC) is the
most sensitive one which shows the intensity and frequency
dependence of the light-induced electron excitations from the
occupied to unoccupied bands \cite{noskov}.

However, the interpretation of the experimental results is
{usually difficult} since {optical data depend over
all possible transitions in the Brillouin zone (BZ) where an}
initial and final states for electron excitations are not known.
On the other hand, {a} comparison of experimental IBOC
spectra with the calculated from the first-principles opens some
perspectives for {a} correct interpretation of experimental
results {if certain transitions are sufficiently strong}.

Optical properties of Ni, Co, and Fe based HA have been widely
investigated theoretically and experimentally
\cite{uba,jpnt,kumar,pico,wan,oehsen}. The most comprehensive
experimental study of various bulk HAs has been {carried
out} by Shreder \emph{et al.}
\cite{shreder1,shreder3,shreder4,shreder5,shreder6,svyazhin}.
However, only a few publications have been focused on the study of
the effect of structural transformations in HA on their optical
properties and hence the electronic structure. Han \emph{et al.}
have shown {significant} influence {of} the alloy
symmetry on the electronic structure and magnetic characteristics
of the Sc-based HA with XA and L2$_{1}$ types of structure
\cite{addi}. Furthermore, Wan \emph{et al.} {have indicated
that the calculated IBOC spectra in a Ni$_{2}$MnGa alloy show a
noticeable structural dependence \cite{wan}. On} the contrary,
Svyazhin \emph{et al.} have experimentally shown that atomic
disorder in bulk Co$_{2}$CrAl HA has no {impact} on its
optical properties \cite{svyazhin}.

Fe$_{2}$MnGa is an interesting example of HA with much richer
structural and magnetic properties than Ni$_{2}$MnGa and a large
ferromagnetic-shape-memory performance {for} slightly
off-stoichiometric compositions \cite{zhu}. Due to structural
instability, Fe-Mn-Ga bulk HA near the 2:1:1 stoichiometric
composition may crystallize in {distinct} crystalline
phases: L1$_{2}$ or/and face-centered cubic (FCC), L2$_{1}$ or/and
body-centered cubic (BCC), mixed BCC+FCC or tetragonal structures
with different types of magnetic order
\cite{zhu,shih,okumura,kudr1}.

The transformations in Fe-Mn-Ga are accompanied by significant
changes {in} their magnetic and transport properties
\cite{zhu,shih,kudr1}. {Such effects give reason to believe
that an electronic band structure also experiences substantial
modifications and hence has an impact on the optical properties of
the alloys}.

{On the other hand, thin films of the corresponding
Fe-Mn-Ga alloys are usually less ordered than the bulk samples and
have a much smaller grain size. Therefore, a comparison of the
optical properties of bulk and film Fe-Mn-Ga alloy samples with
distinct structural ordering and distinct microstructure may, to a
certain extent, assist the verification of the relevance of the
first-principle calculations of the optical properties for the
systems showing structural instability.}

Here we show  the results of first-principles calculations of the
electronic structures and some physical properties of
stoichiometric Fe$_{2}$MnGa alloy with L2$_{1}$ and L1$_{2}$ types
of atomic order. The calculated optical properties were compared
with experimental results obtained for bulk and film samples with
different types of atomic and magnetic orders.

\section{Calculation and experimental details}

{The electronic structures and the IBOC spectra of the
Fe$_{2}$MnGa HA for ordered L2$_{1}$ (225 space group) and
L1$_{2}$ (123 space group) types of structure, as well as
different types of the magnetic order, were calculated by using
the same approach as in our previous article \cite{kudr2}.} The
muffin-tin radii $R_{MT}$ were determined in a way to minimize
inter-sphere volume for the L2$_{1}$ and L1$_{2}$ types of
structure with a smallest their unit-cell volumes. For both types
of the structure $R_{MT}$ were equal to $R_{MT}^{Fe}$ = 2.39 \AA,
$R_{MT}^{Mn}$ = 2.39 \AA and $R_{MT}^{Ga}$ = 2.33 \AA. For
wave-function approximation  of 3$d$-electrons for all the atoms
APW + $l_{0}$ were used. For the wave-functions of other valent
electrons, LAPW basis was employed. To calculate the partial waves
inside MT-spheres maximal orbital quantum numbers equal to $l$=10
and $l$=4 were used. The density plane-wave cutoff was $R_{MT}
K_{max}$ = 7.0. Self-consistency in calculations of the
Fe$_{2}$MnGa alloy IBOC spectra was obtained using 4531 k-points
in the irreducible BZ.

{Two bulk polycrystalline Fe$_{2}$MnGa alloys of different
composition near the stoichiometry 2:1:1 were prepared and
characterized similarly to \cite{kudr1,kudr2}.} 
In addition to bulk samples a set of Fe-Mn-Ga alloy
fine-crystalline films near the stoichiometric composition of
about 50 - 100 nm in thickness was prepared by using DC magnetron
sputtering onto glass and NaCl substrates kept at room temperature
(RT). 
Table \ref{tab1} presents information on the composition and the
structure of fabricated bulk and thin film Fe-Mn-Ga alloy samples.

\begin{table}

\caption{List of investigated bulk and film Fe-Mn-Ga alloy
samples.}\label{tab1}

\begin{tabular}{ccccc}
\hline\hline
          & Sample      &  Heat       &  Sample            & Lattice              \\
Sample No.& composition &  treatment  &  structure         & parameter            \\
          &             &             &                    & (nm)                 \\
\hline\hline

Bulk 1 & Fe$_{49}$Mn$_{25}$Ga$_{26}$ & 723K/150 min & L1$_{2}$  &   0.3717           \\
Bulk 2 & Fe$_{52}$Mn$_{18}$Ga$_{30}$ & 723K/150 min & L2$_{1}$  &   0.5855           \\
Film 1a& Fe$_{56}$Mn$_{20}$Ga$_{24}$ & as-depos.    &  BCC      &   $\sim$ 0.294     \\
Film 1b& Fe$_{56}$Mn$_{20}$Ga$_{24}$ & 673 K/60 min &  FCC      &   $\sim$ 0.377     \\
Film 2a& Fe$_{46}$Mn$_{35}$Ga$_{19}$ & as-depos.    &  BCC      &   $\sim$ 0.306     \\
Film 2b& Fe$_{46}$Mn$_{35}$Ga$_{19}$ & 673 K/60 min &  FCC      &   $\sim$ 0.380     \\
\hline\hline
\end{tabular}
\end{table}

Bulk samples for {the} optical measurements of about 10
$\times$ 20 $\times$ 2 mm$^{3}$ in size were cut from ingots
employing spark erosion technique followed by mechanical polishing
with diamond pasts. To avoid surface contaminations induced by
mechanical polishing bulk samples before optical measurements were
annealed in a high-vacuum condition at $T_{ann}$=723 K for 150
min. Frequency dependence of the optical conductivity (OC)
$\sigma(\hbar\omega)$ was measured by using a spectroscopic
rotating-analyzer ellipsometer in a spectral range of 330 - 2500
nm (3.75 - 0.5 eV) at a fixed incidence angle of 73$^{\circ}$. The
optical properties of bulk Fe-Mn-Ga HA samples were measured at
different temperatures, which provide different types of magnetic
ordering. 
Magnetic properties of bulk and film samples were investigated by
measuring at RT magnetization hysteresis loops employing a
vibrating sample magnetometer and DC magnetic susceptibility in a
temperature range 80 - 750 K.

\section{Results and discussion}

\begin{figure}[tbp]
\textcolor{black}{\includegraphics[width=11cm]{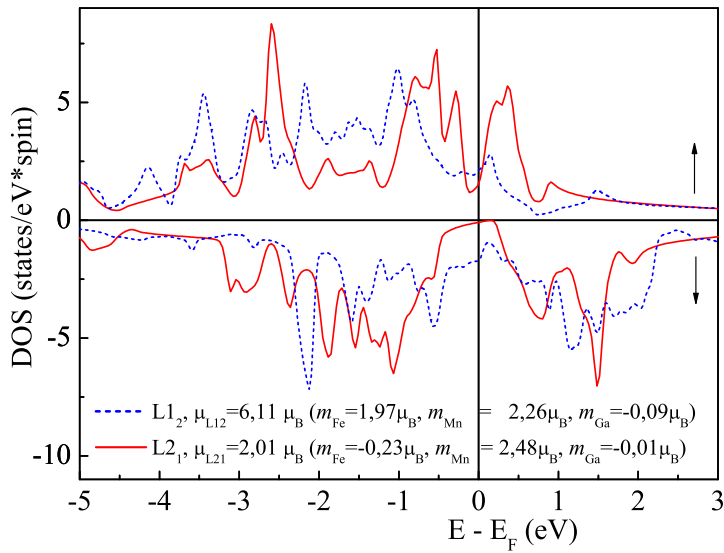}
\caption{Calculated total density of electronic states for
majority ($\uparrow$) and minority ($\downarrow$) spins for the
stoichiometric Fe$_{2}$MnGa alloy  with the L2$_{1}$ type of
structure and an FI type of magnetic ordering and with the
L1$_{2}$ type of structures and an FM type of magnetic ordering.}
\label{fig1}}
\end{figure}

It was found that L2$_{1}$ and L1$_{2}$ types of the structure
with ferrimagnetic (FI) and ferromagnetic (FM) types of magnetic
order, respectively, are most favorable because of the lowest
total electron energy among other types of structure and magnetic
orders \cite{kudr1}. As it is shown in Fig. \ref{fig1}, the
calculated the density of electronic states (DOS) $N(E)$ for these
states are noticeably different. Unlike the case of {the}
L1$_{2}$ type of order, $N(E)$ for the Fe$_{2}$MnGa alloy with the
L2$_{1}$ type of atomic order exhibit pseudo gaps near the Fermi
level both for the majority and minority spins. Because of a
significant difference {in} $N(E)$ for L2$_{1}$ and
L1$_{2}$ structures of the Fe$_{2}$MnGa alloy, one should expect
the corresponding effect on their (calculated) optical properties.

Resulting IBOC spectra in metals are formed by additive
contributions from electron excitations in the majority and
minority electron subbands. IBOC for selected spin bands can be
given as:

\begin{eqnarray}
\sigma(\omega)=\frac{\omega\varepsilon_{2}(\omega)}{4\pi}=\frac{ve^2}{8\pi^2{\hbar}m^2\omega}{\int}d^3k{\sum}_{kk^{'}}[|{\langle}kn|\mathcal{P}|kn^{'}{\rangle}|^{2}{\times}{}\nonumber\\
f(kn)(1-f(kn^{'}))\delta(E_{nk}-E_{n^{'}k}-{\hbar}{\omega})]
\end{eqnarray}

where $\hbar\omega$ is the incident photon energy, $\mathcal{P}$
is the momentum operator,
$\frac{\hbar}{i}\frac{\partial}{{\partial}x^{'}}$, $|kn{\rangle}$
is the eigenfunction with eigenvalue $E_{nk}$, $f(kn)$ is the
Fermi distribution function.

\begin{figure}[tbp]
\textcolor{black}{\includegraphics[width=10cm]{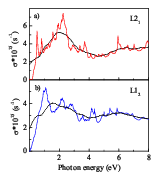}
\caption{Calculated resulting IBOC spectra for stoichiometric
Fe$_{2}$MnGa alloy with a) L2$_{1}$ and b) L1$_{2}$ structures
(solid lines). Dashed lines show corresponding smoothed IBOC
spectra.} \label{fig2}}
\end{figure}

\begin{figure}[tbp]
\textcolor{black}{\includegraphics[width=9cm]{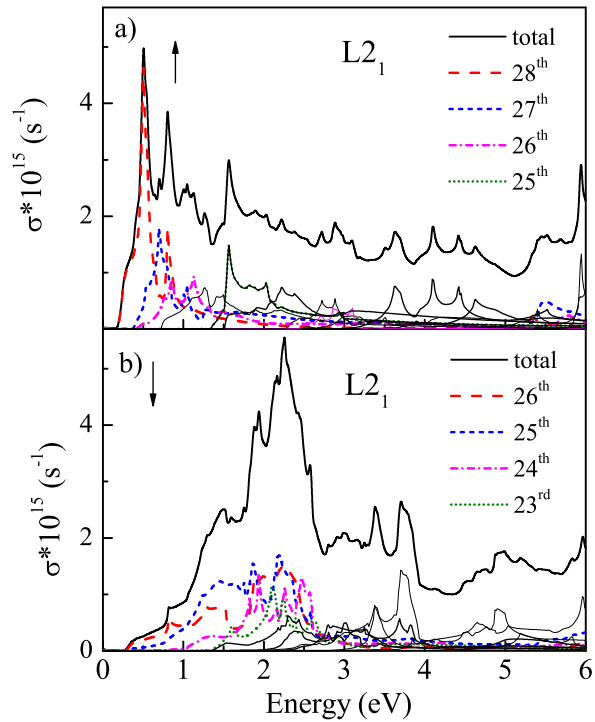}
\caption{Calculated spin-resolved contributions to the IBOC
spectra from the electron excitations in the a) majority
($\uparrow$) and b) minority ($\downarrow$) bands for the
stoichiometric Fe$_{2}$MnGa alloy with the L2$_{1}$ structure.
Numbers indicate the band numbers which are initial for electron
excitations to all possible upper empty bands.} \label{fig3}}
\end{figure}

The calculated resulting and spin-resolved IBOC spectra for
stoichiometric perfectly ordered Fe$_{2}$MnGa HA with L2$_{1}$ and
L1$_{2}$ types of structure are shown in Figs. \ref{fig2},
\ref{fig3} and \ref{fig5}. The resulting IBOC spectrum for the
L2$_{1}$ structure is characterized by the dominant absorption
peak at $\hbar\omega\approx$ 2 eV with less intense peculiarities
(absorption peaks) on its low- and high-energy slopes. For
$\hbar\omega\geq$ 4 eV energy range $\sigma(\hbar\omega)$ for the
L2$_{1}$ phase insignificantly increases with photon energy (see
Fig. \ref{fig2}). The calculated IBOC spectrum for the L1$_{2}$
structure of Fe$_{2}$MnGa HA exhibits a set of less pronounced
absorption peaks located at $\hbar\omega\approx$ 1.0, 2.15 and
2.85 eV, forming together at 0 $<\hbar\omega<$ 3 energy range wide
absorption band. However, its intensity is somewhat smaller than
that for the L2$_{1}$ phase (see Fig. \ref{fig2}). It should be
noted here that the intraband contribution to the calculated OC
spectra of Fe$_{2}$MnGa alloy was not taken into account.

Figure \ref{fig3} shows spin-resolved IBOC spectra of the L2$_{1}$
phase and partial contributions formed by electron excitations
from various occupied bands. It is seen that the most intense
absorption peak at $\hbar\omega\approx$ 2 eV  results from
electron excitations from the minority 23$^{rd}$, 24$^{th}$,
25$^{th}$ and 26$^{th}$ bands. Electron excitations from other
occupied minority bands produce a double peak at
$\hbar\omega\approx$ 3.5 eV located on the energy independent
plateau of the $\sigma(\hbar\omega)$ spectrum for the energy range
of $\hbar\omega>$ 3.0 eV. At the same time, electron excitations
originated from the 28$^{th}$, 27$^{th}$, 26$^{th}$ and 25$^{th}$
majority bands produce a set of narrow and intense IBOC peaks in
the near infra-red energy range whose intensities decrease with
the photon energy.

\begin{figure}[tbp]
\textcolor{black}{\includegraphics[width=12cm]{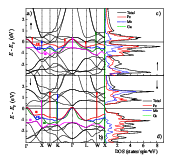}
\caption{Spin-resolved energy band structures [a) and b)] and
element resolved DOS [c) and d)] for stoichiometric Fe$_{2}$MnGa
alloy with L2$_{1}$ type of order for a), c) majority ($\uparrow$)
and b), d) minority ($\downarrow$) spins.} \label{fig4}}
\end{figure}

\begin{figure}[tbp]
\textcolor{black}{\includegraphics[width=10cm]{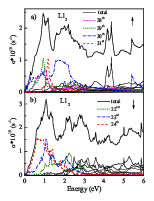}
\caption{Calculated spin-resolved contributions to the IBOC
spectra from the electron excitations in the a) majority
($\uparrow$) and b) minority ($\downarrow$) bands for Fe$_{2}$MnGa
stoichiometric alloy with L1$_{2}$ structure. Numbers indicate the
band numbers which are initial for electron excitations to all
possible upper empty bands.}\label{fig5}}
\end{figure}

Figure \ref{fig4} presents spin-resolved energy band structures
{and element resolved DOS} for stoichiometric Fe$_{2}$MnGa
alloy with the L2$_{1}$ type of atomic order. The bands involved
in the aforementioned excitations as initial are marked by thick
color lines. Vertical arrows with corresponding color show
possible {most intensive} electron excitations which form
partial IBOC peaks.

{It is seen that the most partial interband contributions
to the resulting IBOC spectrum from the definite bands are smeared
over a wide energy range (several eV). Among them, the peak at
$\hbar\omega\approx$ 0.5 eV originating from the 28$^{th}$
majority band probably is the only one with localized energy of
electron excitation (see Fig. \ref{fig3}a). Therefore, for this
case the bands and parts of the Brillouin zone involved in this
peak formation can be found with more certainity. These are the
initial states the near L high symmetry point, the sates along
$\Gamma$ - K and $\Gamma$ - X directions in the BZ located at 0.3
- 0.4 eV below the Fermi level (see Fig. \ref{fig4}a). Figure
\ref{fig4}c clearly shows that these states are related to Fe and
Mn atoms. Their detailed analysis allows us to conclude that they
are mainly $e_{g}$ states of Mn and $d$, $e_{g}$ and $t_{2g}$
states of Fe}.

\begin{figure}[tbp]
\textcolor{black}{\includegraphics[width=12cm]{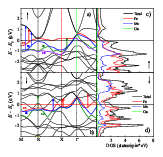}
\caption{Spin-resolved energy band structures [a) and b)] and
element resolved DOS [c) and d)] for stoichiometric Fe$_{2}$MnGa
alloy with L1$_{2}$ type of order for a), c) majority ($\uparrow$)
and b), d) minority ($\downarrow$) spins.} \label{fig6}}
\end{figure}

For the L1$_{2}$ phase of Fe$_{2}$MnGa both spin subbands
contribute nearly equally to the resulting IBOC spectrum formation
(see Fig. \ref{fig5}). It is seen that the electron excitations
initiated from 31$^{st}$, 30$^{th}$, 29$^{th}$ and 28$^{th}$
majority bands produce two intense peaks at $\hbar\omega\approx$ 1
eV and $\hbar\omega\approx$ 2 eV. Electron excitations from other
majority bands produce for the 3 $<\hbar\omega<$ 6 eV energy
region an energy independent plateau with a narrow and intense
peak located at $\hbar\omega\approx$ 4.25 eV. Electron excitations
originated from 22$^{nd}$, 23$^{rd}$ and 24$^{th}$ minority bands
form an intense IBOC peaks at $\hbar\omega\approx$ 1.0, 2.2 and
2.8 eV. Electron excitations from other occupied minority bands
produce a set of low-intense peaks for the 3 $<\hbar\omega<$ 6 eV
energy region. Figure \ref{fig6} presents spin-resolved energy
band structures {and element resolved DOS} for
stoichiometric Fe$_{2}$MnGa alloy with the L1$_{2}$ type of atomic
order. The bands involved in the aforementioned electron
excitations as initial have been marked in this figure by thick
color lines. Vertical arrows with the corresponding color also
show possible {the most intense} electron excitations which
form partial IBOC peaks.

Thus, one can conclude that both calculated IBOC spectra for the
stoichiometric Fe$_{2}$MnGa alloy with L2$_{1}$ and L1$_{2}$ types
of the structure have noticeably different and rather complicated
nature but unfortunately demonstrate some visual similarity in the
spectral shape.

\begin{figure}[tbp]
\textcolor{black}{\includegraphics[width=10cm]{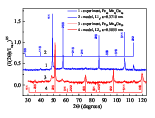}
\caption{Experimental XRD patterns for bulk Fe-Mn-Ga alloys
together with calculated stroke-diagrams for perfectly ordered
stoichiometric Fe$_{2}$MnGa alloy with L2$_{1}$ and L1$_{2}$
orders.} \label{fig7}}
\end{figure}

Figure \ref{fig7} shows  experimental XRD patterns for bulk
Fe-Mn-Ga alloys together with  the simulated stroke-diagrams for
the stoichiometric Fe$_{2}$MnGa alloy with  the L2$_{1}$ and
L1$_{2}$ structures, respectively. Superstructure reflections
[(100) and (110) for Fe$_{49}$Mn$_{25}$Ga$_{26}$ alloy and (200)
and (311) for Fe$_{52}$Mn$_{18}$Ga$_{30}$ alloy] in the
experimental diffraction patterns definitely show on the presence
of L1$_{2}$ and L2$_{1}$ types of atomic order in aforementioned
bulk samples, respectively. At the same time, some presence of the
second phase in these bulk alloy samples should be mentioned. But
the amount of the second phase estimated roughly by the intensity
ratio of the most intense reflections does not exceed several
percents. The coherence length (or {the} mean grain size)
for the bulk Fe$_{52}$Mn$_{18}$Ga$_{30}$ and
Fe$_{49}$Mn$_{25}$Ga$_{26}$ alloys were evaluated by using full
width at half maximum of the XRD peaks and found to be $D$ = 37
and 60 nm, respectively.

\begin{figure}[tbp]
\textcolor{black}{\includegraphics[width=5cm]{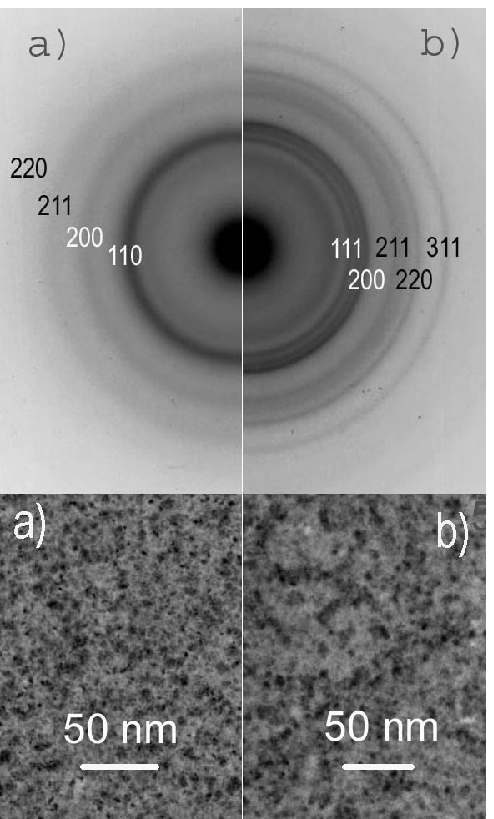}
\caption{Microdiffraction (top panels) and microstructure (bottom
panels) patterns for (a) as-deposited and (b) annealed
Fe$_{46}$Mn$_{35}$Ga$_{19}$ alloy films.}\label{fig8}}
\end{figure}

According to XRD (not shown) and TEM results all RT as-deposited
Fe-Mn-Ga alloy films have fine crystalline (or even
amorphous-like) structure of BCC type (i. e. disordered L2$_{1}$)
with a mean grain size of about $D\approx$ 3 nm. Annealing of the
Fe-Mn-Ga HA films at $T$ = 673 K during 1 hour induces noticeable
changes of the TEM patterns which indicate the formation of a fine
crystalline structure with a short-range-order of FCC type (i. e.
disordered L1$_{2}$) and somewhat larger mean grain size
($D\approx$ 6 nm). As an example, typical for all the investigated
as-deposited and annealed Fe-Mn-Ga alloy films microdiffraction
and microstructure patterns for Fe$_{39}$Mn$_{25}$Ga$_{36}$ alloy
films are shown in Fig. \ref{fig8}.

\begin{figure}[tbp]
\textcolor{black}{\includegraphics[width=10cm]{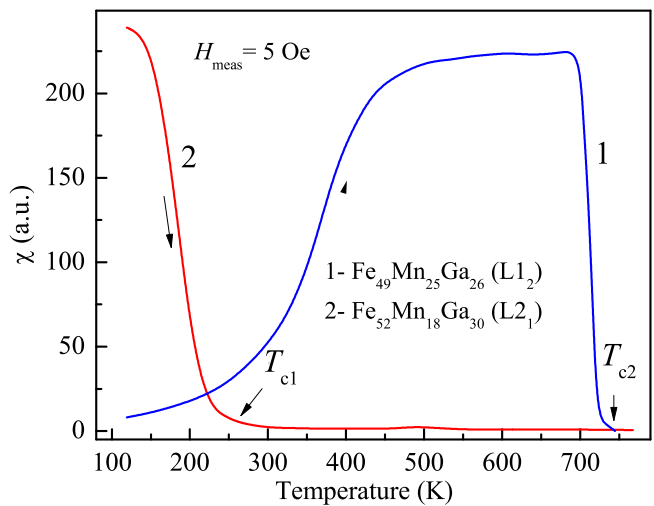}
\caption{Temperature dependencies of magnetic susceptibility for
bulk Fe$_{49}$Mn$_{25}$Ga$_{26}$ (1) and
Fe$_{52}$Mn$_{18}$Ga$_{30}$ (2) alloys.} \label{fig9}}
\end{figure}

\begin{figure}[tbp]
\textcolor{black}{\includegraphics[width=10cm]{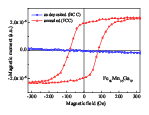}
\caption{In-plane RT magnetization hysteresis loops for
as-deposited and annealed Fe$_{46}$Mn$_{35}$Ga$_{19}$ alloy
films.} \label{fig10}}
\end{figure}

The crystalline structure of bulk and film Fe-Mn-Ga alloy samples
determines their magnetic properties. Thus,
Fe$_{52}$Mn$_{18}$Ga$_{30}$ alloy with the L2$_{1}$ type of
structure behaves as FM [actually FI] with the Curie temperature
$T_{C1}\approx$ 250 K, while Fe$_{49}$Mn$_{25}$Ga$_{26}$ HA with
the L1$_{2}$ type of structure has $T_{C2}\approx$ 750 K. However,
below $T\approx$ 450 K a rapid drop in $\chi(T)$ dependence is
observed for Fe$_{49}$Mn$_{25}$Ga$_{26}$ HA (see  Fig.
\ref{fig9}). Such behavior is usually attributed to metamagnetic
transformation from an FM to an antiferromagnetic (AFM) state
\cite{tang}. At the same time induced by annealing the transition
from a BCC to an FCC type of structure in Fe-Mn-Ga alloy films
causes their transition from a paramagnetic to an FM state (see
Fig. \ref{fig10}).

For a correct comparison of the calculated IBOC spectra with
experimental ${\sigma(\hbar\omega})$ spectra, the intraband
contributions should be {subtracted} from the experimental
OC spectra. Accurately it can be done by using the measurement
results obtained in far infra-red region ($\lambda \geq$ 10
$\mu$m) where intraband (Drude) contribution dominates over the
interband term. The optical properties in the intraband absorption
region can be explained by Drude formula:
\begin{eqnarray}
\varepsilon_{1}=1-\frac{\Omega^2_{p}}{\omega^2+\gamma^2}
\end{eqnarray}

\begin{eqnarray}
4\pi\sigma=\frac{\Omega^2_{p}\gamma}{\omega^2+\gamma^2}
\end{eqnarray},

where $\omega$ is light frequency, $\varepsilon_{1}$, $\Omega_{p}$
and $\gamma$ are real part of the dielectric function, plasma and
effective relaxation frequencies of free charge carriers,
respectively. $\Omega_{p}$ and $\gamma$ can be determined by
plotting so-called Argand diagrams:

\begin{eqnarray}
\frac{1}{1-\varepsilon_{1}}=\frac{\gamma^2}{\Omega^2_{p}}+\frac{1}{\Omega^2_{p}}\times\omega^2
\end{eqnarray}

The linear part of the 1/(1-$\varepsilon_{1}$) dependence plotted
in the $\omega^2$ scale evidences on the dominance of the Drude
absorption in the optical properties of the sample. Having
$\Omega$ and $\gamma$ values determined in such a way, the
intraband conductivity can be calculated. Despite the fact that
our optical measurements were restricted by 2.5 $\mu$m edge, the
Argand diagrams for all the investigated samples in the 1.0 $<
\lambda <$ 2.5 $\mu$m spectral range were {almost} linear.

Furthermore, to simulate temperature, structurally and chemically
induced imperfections of the real samples the calculated IBOC
spectra were smoothed (see Fig. \ref{fig2}).

\begin{figure}[tbp]
\textcolor{black}{\includegraphics[width=10cm]{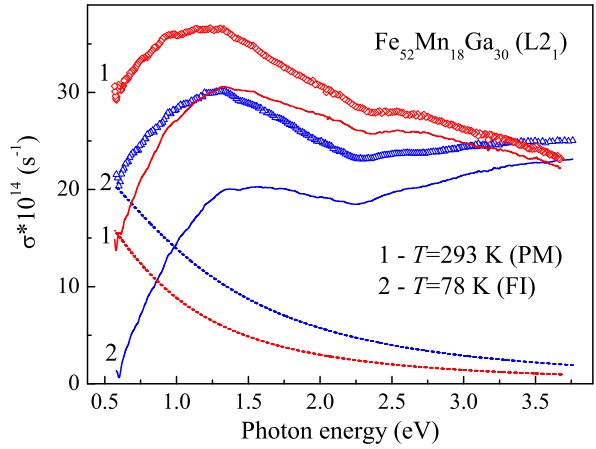}
\caption{Experimental $\sigma(\hbar\omega)$ spectra for
{the} Fe$_{52}$Mn$_{18}$Ga$_{30}$ alloy with the L2$_{1}$
structure taken at different temperatures and different types of
the magnetic ordering (symbols) together with corresponding
extracted intra - (dashed lines) and interband (solid lines)
contributions.}\label{fig11}}
\end{figure}

\begin{figure}[tbp]
\textcolor{black}{\includegraphics[width=10cm]{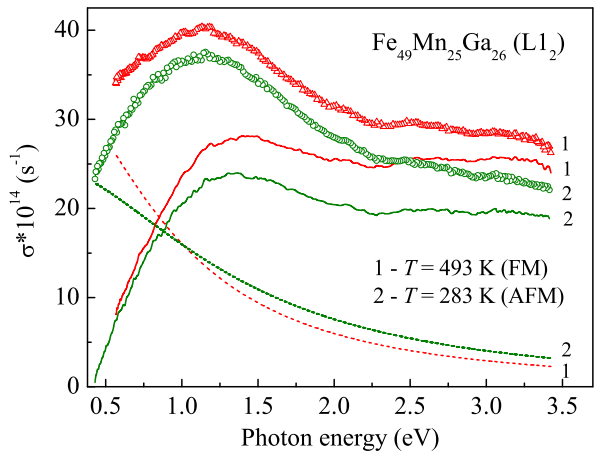}
\caption{Experimental $\sigma(\hbar\omega)$ spectra for
{the} Fe$_{49}$Mn$_{25}$Ga$_{26}$ alloy with the L1$_{2}$
structure taken at different temperatures and different types of
the magnetic ordering (symbols) together with corresponding
extracted intra - (dashed lines) and interband (solid lines)
contributions.}\label{fig12}}
\end{figure}

The experimental $\sigma(\hbar\omega)$ spectra for {the}
bulk magnetically ordered (FI) Fe$_{52}$Mn$_{18}$Ga$_{30}$ HA with
the L2$_{1}$ type of structure ($T_{meas}$ = 78 K) and
{the} FM ordered Fe$_{49}$Mn$_{25}$Ga$_{26}$ HA with the
L1$_{2}$ structure ($T_{meas}$ = 493 K) are shown in Figs.
\ref{fig11}  and \ref{fig12}.

It should be mentioned here that the experimental interband
$\sigma_{L21}^{FI}(\hbar\omega)$ spectrum {corresponds} in
spectral shape {to the} smoothed calculated IBOC spectrum
(see Figs. \ref{fig2} and \ref{fig11}) and is characterized by the
presence of a broad interband absorption peak located at
$\hbar\omega\approx$ 1.5 eV. Above $\hbar\omega\approx$ 2.25 eV
some growth of $\sigma$ indicates an increased interband
absorption in the alloy (see Fig. \ref{fig11}). Thus, a reasonable
qualitative correspondence between the shapes of the calculated
and experimental OC spectra for the magnetically ordered L2$_{1}$
phase of Fe-Mn-Ga alloy proves the sufficient quality of the IBOC
calculations and allows us to single out the peak at
$\hbar\omega\approx$ 1.5 eV as the electron excitations between
aforementioned states (see Figs. \ref{fig2} and \ref{fig11}). At
the same time, the lack of fine structures in the experimental
$\sigma(\hbar\omega)$ spectrum which are visible on the calculated
spectrum can be explained by a polycrystalline microstructure of
the bulk sample.

The correspondence between the shapes of the calculated and
experimental interband $\sigma(\hbar\omega)^{FM}_{L12}$ spectra
for the FM ordered Fe$_{2}$MnGa HA with the L1$_{2}$ structure is
also rather good: the experimental interband $\sigma(\hbar\omega)$
spectrum exhibits {a} less pronounced and somewhat
red-shifted absorption peak in comparison with that for
{the} L2$_{1}$ phase. Good correspondence between
calculated and experimentally determined IBOC spectra for L2$_{1}$
and L1$_{2}$ phases should be mentioned (see Figs. \ref{fig2},
\ref{fig11} and \ref{fig12}).

The optical properties of bulk electrolytically polished Fe-Mn-Ga
alloys were investigated earlier by Kr$\acute{a}$l \cite{kral}. It
should be noted that all the investigated by him alloys of
different compositions (and probably of different crystalline
structures) demonstrate rather similar in spectral shape and
absolute value OC spectra \cite{kral}.

\begin{figure}[tbp]
\textcolor{black}{\includegraphics[width=10cm]{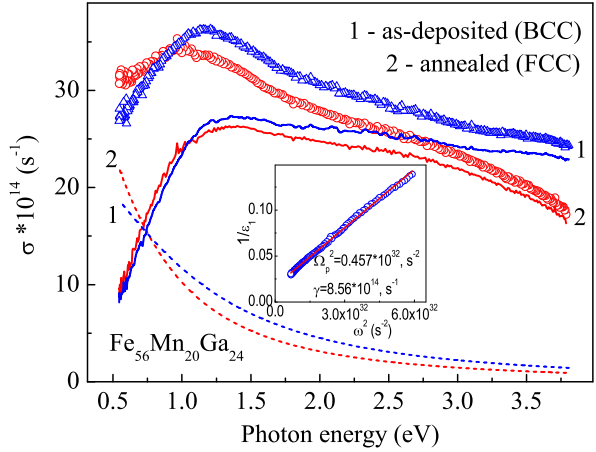} \caption{RT
experimental OC spectra of the as-deposited and annealed
Fe$_{56}$Mn$_{20}$Ga$_{24}$ alloy film (symbols) together with
extracted corresponding intra- (dashed lines) and interband (solid
lines) contributions. Inset shows an example of the Argand diagram
for annealed state.}\label{fig13}}
\end{figure}

\begin{figure}[tbp]
\textcolor{black}{\includegraphics[width=10cm]{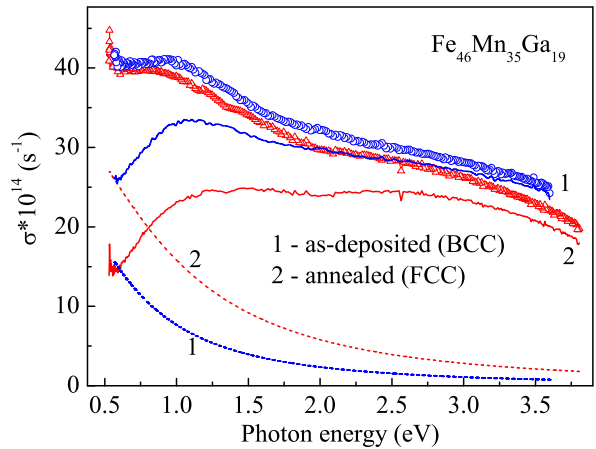} \caption{RT
experimental OC spectra of the as-deposited and annealed
Fe$_{46}$Mn$_{35}$Ga$_{19}$ alloy film (symbols) together with
extracted corresponding intra- (dashed lines) and interband (solid
lines) contributions.}\label{fig14}}
\end{figure}

Thin film Fe-Mn-Ga samples are usually of much worse structural
quality than the corresponding bulk samples (a smaller mean grain
size, lower degree of atomic order, etc.). However, the OC spectra
for the Fe-Mn-Ga films with BCC (i. e. the disordered L2$_{1}$
phase) and FCC (i. e. the disordered L1$_{2}$ phase) structures
look rather similar to the corresponding $\sigma(\hbar\omega)$
spectra of bulk samples (see Figs.\ref{fig11} - \ref{fig14}).
Therefore, the L2$_{1} \rightarrow$ BCC and the L1$_{2}
\rightarrow$ FCC {types of the} atomic disorder do not
change significantly the electronic states responsible for main
interband peaks formation.

The type of magnetic ordering has no impact on the experimental
optical properties and hence on electronic structures of bulk
Fe$_{52}$Mn$_{18}$Ga$_{30}$ and Fe$_{49}$Mn$_{25}$Ga$_{26}$ alloys
(see Figs. \ref{fig11} and \ref{fig12}). This may be due to the
so-called specific time scale of forming the electronic structure
and spin fluctuations. It is known that the time needed to
establish the electron energy spectrum is determined by the
lifetime of the excited states defined as $t_{es}\sim
\hbar/W_{d}\sim$ 10$^{-15}$ s (where $W_{d}$ is the
$d$-bandwidth). At the same time, collective electrons need much
more time to create spin correlations:
$t_{sc}\sim\hbar/k_{B}T_{C}\approx$ 10$^{-13}$ s \cite{vons}.
Because of a large number of the electrons involved in spin
fluctuations, $t_{sc} \gg t_{es}$, and one-particle states have
enough time to build-up in the relatively slowly changing
potentials of spin fluctuations. Thus, the local electronic
structure is formed in slowly fluctuating (nearly constant for
electron jumps) local field. The only difference of this local
electronic structure from that of obtained at $T$= 0 is the random
orientation of quantization axis \cite{vons}.

\section{Summary}

Electronic structure [DOS, E(k), magnetic moment] and IBOC for
stoichiometric Fe$_{2}$MnGa alloy with the L2$_{1}$ and L1$_{2}$
structures have been calculated.

It was shown that the calculated IBOC spectra for these phases of
Fe$_{2}$MnGa alloy have rather complicated nature - a large number
of the majority and minority bands are involved in electron
excitations, {which form} the resulting IBOC spectra.

The optical properties of bulk Fe-Mn-Ga alloys near the
stoichiometry 2:1:1 with the L2$_{1}$ and L1$_{2}$ types of
structure have been experimentally measured. Correspondence in
spectral shape between the experimental and calculated IBOC
spectra for the L2$_{1}$ phase of Fe-Mn-Ga alloy allows us to
correlate the peaks observed in the experiment with an electron
excitations between corresponding electronic bands.

Structural L2$_{1}$ $\rightarrow$ L1$_{2}$ or BCC $\rightarrow$
FCC transformations in bulk and film Fe-Mn-Ga alloy samples causes
drastic changes in their magnetic properties. At the same time,
magnetic transformations in bulk Fe$_{52}$Mn$_{18}$Ga$_{30}$ and
Fe$_{49}$Mn$_{25}$Ga$_{26}$ alloys do not {affect} their
IBOC spectra and hence the electronic states responsible for these
spectra formation. We associate  such the independence of the
optical properties {on the magnetic ordering} of Fe-Mn-Ga
alloys with the specific time scale of forming the electronic
structure and spin fluctuations in alloys.

{Little impact of structural ordering of Fe-Mn-Ga alloys on
their optical properties suggests that the main features of the
electronic structure of the alloys are mainly determined by their
short-range order.}

\ack

This work has been supported by the project ``Marie
Sk{\l}odowska-Curie Research and Innovation Staff Exchange
(RISE)'' Contract No. 644348 with the European Commission, as part
of the Horizon2020 Programme and by the project No. 84/18-H of the
program "Fundamental problems of the creation of new nanomaterials
and nanotechnologies". 
Authors appreciate  A. E. Perekos, A.
V. Terukov and V. I. Bodnarchuk for their help in the $\chi(T)$, XRD and TEM
measurements.


\section*{References}

\begin{thebibliography}{99}

\bibitem{graf2011}Graf T, Felser C and Parkin S 2011 {\it Progress in solid state chemistry} {\bf39} 1

\bibitem{uba}Uba S, Bonda A, Uba L, Bekenov L V, Antonov V N and Ernst A 2016 {\it Phys. Rev.} B {\bf94} 054427

\bibitem{noskov}Noskov M M 1983 {\it Optical and magneto-optical
properties of metals}, (Sverdlovsk, UNC AN SSSR) p 220

\bibitem{jpnt} Kubo Y, Takakura N and Ishida S 1983 {\it J. Phys. F: Met. Phys.} {\bf13} 161

\bibitem{kumar} Kumar M, Nautiyal T and Auluck S 2009 {\it J. Phys.
: Condens. Matter} {\bf21} 196003

\bibitem{pico} Picozzi S, Continenza A and Freeman A J 2006 {\it J.
Phys. D: Appl. Phys.} {\bf39} 851

\bibitem{wan}Wan J F and Wang J N 2005 {\it Physica} B, {\bf355} 172

\bibitem{oehsen} von Oehsen S, Scholtyssek S M, Pels C,
Neuber G, Rauer R, R\"{u}bhausen M and Meier G 2005 {\it Jour.
Magn. Magn. Mater.} {\bf290-291} 1371

\bibitem{shreder1} Shreder E I, Svyazhin A D and Fomina K A 2012
{\it Physics of Metals and Metallography} {\bf113} 146

\bibitem{shreder3}Shreder E I, Kirillova M M  and Dyakina B P 1996
{\it Physics of Metals and Metallography} {\bf81} 406

\bibitem{shreder4} Shreder E I, Streltsov S V, Svyazhin A, Makhnev A, Marchenkov V V, Lukoyanov A
and Weber H W 2008 {\it J. Phys.: Cond. Matter} {\bf20} 045212

\bibitem{shreder5} Shreder E I, Kirillova M M and Dyakina B P 2000
{\it Physics of Metals and Metallography} {\bf90} 362

\bibitem{shreder6} Fomina K A, Marchenkov V V, Shreder E I and
Weber H W 2011 {\it Solid State Phenomena} {\bf168-169} 545

\bibitem{svyazhin} Svyazhin A D, Shreder E I, Voronin V I,
Berger I F and Danilov S E 2013 {\it Jour. Exp. Theor. Phys.}
{\bf143} 518 (in Russian)

\bibitem{addi}Han Y, Chen Z, Kuang M, Liu Z, Wang X, Wang X 2019, Results in Physics, {\bf12}, 435

\bibitem{zhu} Zhu W, Liu E K, Feng L, Tang X D, Chen J L, Wu G H,
Liu H Y, Meng F B and Luo H Z 2009 {\it Appl. Phys. Lett.} {\bf95}
222512

\bibitem{shih}Shih C W, Zhao X G, Chang H W, Chang W C and Zhang Z D 2013 {\it Jour. All. Comp.}
{\bf570} 14

\bibitem{okumura} Okumura H, Hashiba E and Nagata K 2014
{\it Intermetallics} {\bf49}, 65

\bibitem{kudr1}Kudryavtsev Y V, Perekos A E, Uvarov N V, Kolchiba M R,
Synoradzki K and Dubowik J 2016 {\it Jour. Appl. Phys.} {\bf119}
205103

\bibitem{kudr2}Kudryavtsev Y V, Uvarov V N, Iermolenko V N,
Glavatskyy I N, Dubowik J, 2012 {\it Acta Materialia} {\bf60} 4780

\bibitem{blaha}Blaha P, Schwarz K,  Madsen G K,  Kvasnicka H D and
Luitz L 2001 {\it WIEN2k, An Augmented Plane Wave + Local Orbitals
Program for Calculating Crystal Properties} (Karl-heinz Schwarz,
Techn. Universit Aat Wien, Wien, Austria)

\bibitem{wimmer} Wimmer E, Krakauer H, Weinert M and Freeman A J  1981 {\it Phys.
Rev}. B {\bf24} 864

\bibitem{perdew}Perdew J P,  Burke K and Ernzerhof M 1996 {\it Phys. Rev. Lett.} {\bf77} 3865

\bibitem{uvarov}Uvarov N V, Kudryavtsev Y V, Kravets A F, Vovk A Ya,
Borges R P, Godinho  M and Korenivski V 2012 {\it Jour. Appl.
Phys}. {\bf112} 063909

\bibitem{miura}Miura Y, Nagao K and Shirai M 2004 {\it Phys. Rev}. B,
{\bf69} 144413

\bibitem{tang}Tang X D, Wang W H, Zhu W, Liu E K, Wu G H,
Meng F B, Liu H Y and Luo H Z 2010 {\it Appl. Phys. Lett}. {\bf97}
242513

\bibitem{kral}Kr$\acute{a}$l D 2017 \emph{Optical and magneto-optical properties of Heusler compounds} (Master thesis, Institute of Physics of
Charles University, Prague)

\bibitem{vons}Vonsovskij S V 1986 in \emph{Dynamic and kinetic properties of magnetics},
(edited by Vonsovskij S V and Turov E A, Science publisher, Moscow
248 p, in Russian).

\end{thebibliography}
\end{document}